\documentclass[conference]{IEEEtran}
\usepackage{amssymb}
\usepackage{mathbbold}
\usepackage{amsmath}
\usepackage{amsfonts}
\usepackage{graphicx,cite,epsfig,amssymb,amsmath,epstopdf,algorithm,algorithmic,subfigure}
\usepackage{hyperref}
\usepackage{graphicx,epstopdf}

\newcommand{\be}{\begin{equation}}
\newcommand{\ee}{\end{equation}}
\newcommand{\bee}{\begin{eqnarray}}
\newcommand{\eee}{\end{eqnarray}}
\newcommand{\nnb}{\nonumber}

\newtheorem{pro}{\textbf{Proposition}}

\newtheorem{theo}{\textbf{Theorem}}

\begin{document}

\title{Utility Maximization for Uplink MU-MIMO: Combining Spectral-Energy Efficiency and Fairness \\ Technical Report}

\author{\IEEEauthorblockN{Lei Deng\IEEEauthorrefmark{1}, Wenjie Zhang\IEEEauthorrefmark{2}, Yun Rui\IEEEauthorrefmark{3},
Yeo Chai Kiat\IEEEauthorrefmark{4}}
\IEEEauthorblockA{\IEEEauthorrefmark{1}Department of Information Engineering, The Chinese University of Hong Kong, Hong Kong}
\IEEEauthorblockA{\IEEEauthorrefmark{2} School of Computer Science, Minnan Normal University, China}
\IEEEauthorblockA{\IEEEauthorrefmark{3}Shanghai Advanced Research Institute, Chinese Academic of Science, Shanghai, China}
\IEEEauthorblockA{\IEEEauthorrefmark{4} School of Computer Engineering, Nanyang Technological University, Singapore}
Email:  dl013@ie.cuhk.edu.hk,  zhan0300@ntu.edu.sg, ruiy@sari.ac.cn, asckyeo@ntu.edu.sg}
% make the title area
\maketitle

% make the title area
\maketitle

% As a general rule, do not put math, special symbols or citations
% in the abstract or keywords.
\begin{abstract}
Driven by green communications, energy efficiency (EE) has become a new important criterion for designing wireless communication systems. However, high EE often leads  to low spectral efficiency (SE), which spurs the research on EE-SE tradeoff.  In this paper, we focus on how to maximize the utility in physical layer for an uplink multi-user multiple-input multiple-output (MU-MIMO) system, where we will not only consider EE-SE tradeoff in a unified way, but also ensure user fairness. We first formulate the utility maximization problem, but it turns out to be non-convex. By exploiting the structure of this problem, we find a convexization procedure to convert the original non-convex problem into an equivalent convex problem, which has the same global optimum with the original problem. Following the convexization procedure, we present a centralized algorithm to solve the utility maximization problem, but it requires the global information of all users. Thus we propose a primal-dual distributed algorithm which does not need global information and just consumes a small amount of overhead. Furthermore, we have proved that the distributed algorithm can converge to the global optimum. Finally, the numerical results show that our approach can both capture user diversity for EE-SE tradeoff and ensure user fairness, and they also validate the effectiveness of our primal-dual distributed algorithm.
\end{abstract}

% Note that keywords are not normally used for peerreview papers.

\section{Introduction}
Among the total worldwide energy consumption, communication networks have contributed increasingly from 1.3\% in 2007 to 1.8\% in 2012, and this proportion is anticipated to grow continuously in the coming years \cite{trend12}. This stimulates the fast development of \emph{green  communications} recently \cite{Richard12}. Compared to spectral efficiency (SE), \textit{energy efficiency} (EE), defined as the number of bits that can be transmitted with per energy consumption, becomes a new important criterion for designing green wireless systems. How to obtain optimal EE has become a hot research topic in different wireless communication systems \cite{Feng13}.

On the other hand, multiple-input multiple-output (MIMO) has been a key technique in modern wireless communication systems, because it can significantly increase SE by exploiting transmit diversity and spatial multiplexing gains \cite{Tse05}. MIMO system is used for one single transmitter and one single receiver in a point-to-point way, so it is often referred to single-user MIMO (SU-MIMO). However, in some applications, especially in cellular networks, it is often difficult to install many antennas due to the size limitations of many devices such as smartphones and tablets. To increase the network-wide SE, multi-user MIMO (MU-MIMO) technique has been proposed. Although distributed users only have a small number of antennas or even just one, they can share the same time-frequency resource block to form a MU-MIMO system\cite{3GPP-TR}. In this paper, we are interested in the uplink MU-MIMO because users, such as smartphones and tablets, are often more energy-sensitive.

%Very often, MU-MIMO system consists of many distributed users and one Node-B which deploys multiple antennas. In this scenario, MU-MIMO system has one downlink from Node-B to users, named MIMO broadcast channels (MIMO BC) and one uplink from users to Node-B, named MIMO multiple access channels (MIMO MAC). In this paper, we are more interested in the uplink MU-MIMO because users, such as smartphones and tablets, are often more energy-sensitive.

Recently there are some papers studying how to maximize EE for uplink MU-MIMO system. In \cite{Miao13}, Miao investigates the uplink MU-MIMO system where each user deploys multi-antennas and he demonstrates that EE is maximized when some antennas are turned off if the corresponding  spatial channel is not good or the corresponding  circuit power consumption is large. In \cite{RZ13}, Rui et al. study the uplink MU-MIMO system where each user deploys only one antenna and they maximize EE by jointly doing mode selection and optimal power allocation.

However,  it is well-known that SE and EE are two conflicting objectives  \cite{CS11}. Often high EE leads to low SE and vice verse, which means it is more practical to consider SE and EE simultaneously. Thus how to study the EE-SE tradeoff has attracted a lot of attention \cite{Heliot12, Xiong11, He12, HJ13, DR13} whereas, only a few articles study uplink MU-MIMO system. The authors in \cite{HJ13} consider how to get the  EE-SE tradeoff for a large-scale uplink MU-MIMO system in a system level. They study EE-SE tradeoff in low and high SE regime asymptotically and do not involve user fairness explicitly which is important in multi-user system. Different from \cite{HJ13}, our paper investigates uplink MU-MIMO system in the link level rather than the system level. More importantly, we study EE-SE tradeoff in a unified way and we also guarantee fairness among users. Specifically, our contributions are three-fold,

\begin{itemize}
\item We construct a utility function of all users which not only captures the user diversity for EE-SE tradeoff in a unified way, similar to \cite{DR13}, but also guarantees fairness among all users. Then we maximize the utility function. To the best of our knowledge, we are the first to study EE-SE tradeoff and user fairness together in uplink MU-MIMO system.
\item Although the original utility maximization problem is not convex, we have proposed an approach to convert it into an equivalent convex programming problem which has the same optimal solution with the original problem. This convexization procedure also generates our optimal power allocation scheme in a centralized manner.
\item Apart from the centralized algorithm, we further devise a primal-dual distributed algorithm which only consumes a small amount of overhead between each user and Node-B. Moreover, we have proved that the distributed algorithm converges to the global optimal solution.
\end{itemize}

The rest of this paper is outlined as follows. We describe the system model and formulate the problem in Section II. In Section III, we analyze the optimal power allocation by converting the optimization problem into a convex programming problem. Next in Section IV, we propose a primal-dual distributed algorithm, which can achieve the global optimum. The numerical results are shown in Section V, followed by conclusion in Section VI. Throughout this paper, we
will use $[\cdot]_{ij}$ to denote the matrix's entry in $i$-th row and $j$-th column, $\mathbf{E}[\cdot]$ to denote expectation, $\mathbf{I}_n$ to denote the $n \times n$ identity matrix, and the superscript $\dag$ to denote Hermitian transpose.

\section{System Model and Problem Formulation}
\subsection{System Model}
\begin{figure}[h!]
\centering
  % Requires \usepackage{graphicx}
  \includegraphics[width=0.7\linewidth]{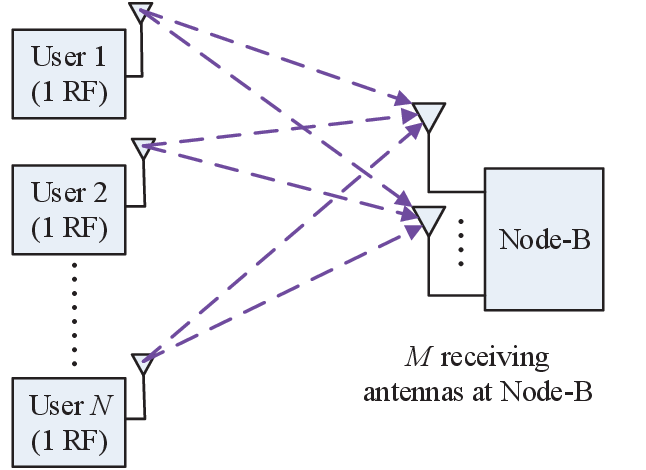}\\
  \caption{System Model}\label{fig_system_model}
\end{figure}

Consider a MU-MIMO system with $N$ users indexed from $1$ to $N$, and one Node-B in a single cell. In this paper, we assume that each user is only equipped with one transmitting antenna, and the Node-B is equipped with $M (M \ge N)$ receiving antennas, as shown in Fig. \ref{fig_system_model}. In uplink, all $N$ users share the same time-frequency resource to transmit data to the Node-B. Denote $P_i$ as the transmit power for user $i$. Then the received signal vector $\mathbf{y} \in \mathbb{C}^{M \times 1}$ is,
\be
\mathbf{y} =\mathbf{H} \mathbf{s} + \mathbf{n},
\ee
where $\mathbf{s} \in \mathbb{C}^{N \times 1}$ denotes the transmit signal vector with ${\mathbf{E}}[{\mathbf{ss}}^\dag ] = \mathrm{diag}\{{P_1}, P_2, \cdots ,{P_N} \}$,
%$\mathbf{G}^{\frac{1}{2}}=\mathrm{diag}\{\sqrt{G_1}, \sqrt{G_2} , \cdots , \sqrt{G_N}\}$ denotes the path-loss matrix, $\mathbf{P}^{\frac{1}{2}}=\mathrm{diag}\{\sqrt{P_1}, \cdots ,\sqrt{P_N} \}$ denotes
%the power loading matrix,
$\mathbf{H} \in \mathbb{C}^{M \times N}$ denotes the channel matrix, and ${\mathbf{n}} \in \mathbb{C}^{M \times 1}$ denotes the additive white Gaussian noise (AWGN) with zero mean and covariance matrix ${\mathbf{E}}[{\mathbf{nn}}^\dag ] = \sigma_n^2 {\mathbf{I}}_M $.

In this paper, we assume that Node-B has perfect channel state information (CSI) for all users and the receiver at Node-B uses zero forcing (ZF) detection method. Thus, the decoded signal vector is
\be
\mathbf{H}^{\#} \mathbf{y} = \mathbf{s} + \mathbf{H}^{\#} \mathbf{n},
\ee
where $\mathbf{H}^{\#} = (\mathbf{H}^\dag \mathbf{H})^{-1} \mathbf{H}^\dag$ denotes the pseudo-inverse of channel matrix $\mathbf{H}$.
Then the signal-to-interference-plus-noise ratio (SINR) at the Node-B's receiver for user $i$ is,

\begin{equation}
%\gamma_i = \underbrace{\frac{P_iG_i}{\sigma_n^2}}_{\rho_i} \underbrace{\frac{1}{[(\mathbf{H}^\dag \mathbf{H})^{-1}]_{ii}}}_{\xi_i}
\gamma_i = \frac{P_i}{\sigma_n^2 [(\mathbf{H}^\dag \mathbf{H})^{-1}]_{ii}} =  P_i \underbrace{\frac{1}{\sigma_n^2 [(\mathbf{H}^\dag \mathbf{H})^{-1}]_{ii}}}_{\delta_i}.
\end{equation}

Then we can obtain SE and EE for user $i$ as
%\be
%SE_i = \log(1+\gamma_i) = \log(1+\delta_i P_i),
%%\quad \text{bps/Hz},
%\ee
%\be
%\label{equ:eei}
%EE_i = \frac{SE_i}{P_i + P^c_i} = \frac{\log(1+\delta_i P_i)}{P_i + P^c_i},
%%\quad \text{bps/Hz/Watts},
%\ee
\bee
&& SE_i = \log(1+\gamma_i) = \log(1+\delta_i P_i), \\
&& EE_i = \frac{SE_i}{P_i + P^c_i} = \frac{\log(1+\delta_i P_i)}{P_i + P^c_i}, \label{equ:eei}
\eee
where $P^c_i$ is a positive constant circuit power consumed by the relevant electronic devices for user $i$.

\subsection{Problem Formulation}
Next, we will construct the utility function in two steps. First, we consider the EE-SE tradeoff. Inspired by the widely-used Cobb-Douglas production function in economics \cite{CD28}, we adopt this model \textit{empirically} to get the ``production" of SE and EE for user $i$,
\be
u_i=(SE_i)^{w_i}(EE_i)^{1-w_i},
\label{equ_uui}
\ee
where $w_i \in [0, 1]$. More specifically, we can regard $(w_i, 1-w_i)$ as a priori articulation of preferences for SE and EE, which captures EE-SE tradeoff in a unified way \cite{DR13}.

Second, we consider the fairness among all $N$ users. If we apply the \textit{proportional fairness} metric, we can define the final utility function for user $i$ as
\bee
U_i(\mathbf{P}) &=& \log(u_i) = \log[(SE_i)^{w_i}(EE_i)^{1-w_i}] \nonumber \\
&=& \log[\log(1+\delta_i P_i)] - (1-w_i)\log(P_i+P_i^c) \nonumber \\
&=& U_i(P_i), \label{equ_ui}\eee
where $\mathbf{P}=(P_1, P_2, \cdots, P_N)$ and the last step shows that the utility for user $i$ is not related to the transmit power of other users.

Based on the utility function in (\ref{equ_ui}), we then formulate our utility maximization problem subject to a power constraint for each user and a power sum constraint for all users,
\bee
   \text{maximize} && \sum_{i=1}^{N} U_i(P_i) \label{equ_object} \\
(\mathbf{P1})  \quad \text{subject to}  &&  \label{equ_cons1} 0 \le P_i \le P^{\max}_i,  \quad \forall i\\
&& \label{equ_cons2} \sum_{i=1}^{N} P_i \le P_{\max}.
\eee
In (\ref{equ_object}), we aim at maximizing the sum of the utility for all users, i.e., the network-wide utility. Inequality (\ref{equ_cons1}) is the individual power constraints where $P^{\max}_i$ is the maximal transmit power for user $i$. Inequality (\ref{equ_cons2}) is the power sum constraint for the total MU-MIMO system where $P_{\max}$ is the maximal transmit power for all users, which is the \textit{power budget} of the whole system.
%The reason why we set $P_{\max}$ is as follows. From the perspective of the whole uplink MU-MIMO system or Node-B, the whole system performance is more important than individual utility. Thus the authors in \cite{Guo14} try to minimize the total power consumption of all users while meeting some pre-defined SINR targets. However, we capture the same idea in a reversed way, namely, we aim at achieving the best system performance given a maximal power sum constraint $P_{\max}$ for all users.
%In addition, by setting $P_{\max} = \infty$, our formula in (\ref{equ_cons2}) also includes the scenario without any power budget limitation.

\section{Optimal Power Allocation}
In the previous section, we have formulated the problem to maximize the network-wide utility in (\ref{equ_object}), which however is not a concave function  since  EE in (\ref{equ:eei}) is neither convex nor concave\cite{DR13}. Therefore, in this section, we will exploit the inner structure of ($\mathbf{P1}$) and find that we can narrow down the feasible region without changing the global optimum. Furthermore, we will prove that the objective function in (\ref{equ_object}) is concave in the new feasible region, which converts the original problem into a convex programming problem. After some analysis, we will get the optimal power allocation scheme for $(\mathbf{P1})$ with a centralized algorithm.

\subsection{Convexization Procedure}
To narrow down the feasible region in ($\mathbf{P1}$), we first consider the individual power constraints in (\ref{equ_cons1}). Since the optimization problem can be changed as,
\be
\max_{0 \le P_i \le P^{\max}_i, \forall i } \sum_{i=1}^{N} U_i(P_i) = \sum_{i=1}^{N} \max_{0 \le P_i \le P^{\max}_i} U_i(P_i),
\ee
we just need to find the maximal individual utility, i.e., $U_i(P_i)$ for any user $i \in \{1,2,\cdots, N\}$. For the individual utility function $U_i(P_i)$ in (\ref{equ_ui}),
%\be
%\label{ui_u}
%U_i(P_i) = \log\alpha_i + \log(\log(1+\delta_i P_i)) - (1-w_i)\log(P_i+P_i^c),
%\ee
we have the following proposition.

\begin{pro}
  \label{pro1}
For any user $i$ under individual power constraint in (\ref{equ_cons1}), there exists one and only one point $P^u_i\in(0, P_i^{\max}]$ that maximizes $U_i (P_i )$. The function $U_i(P_i)$ is strictly increasing and strictly concave over the interval $[0, P^u_i]$
 while strictly decreasing over the interval $(P^u_i, P_i^{\max}]$.
 In addition, $P_i^u$ can be derived as follows,
 \be
 P_i^u =
 \begin{cases}
 P_i^{\max} & \text{if} \quad w_i > 1 - \beta(P_i^{\max}) \\
 P_i^0  & \text{if} \quad w_i \le 1-\beta(P_i^{\max})
 \end{cases}
 \ee
 where
 \be
\label{equ1}
\beta(P_i)=\frac{\delta_i(P_i + P_i^c)}{(1 + \delta_i P_i)\log(1 + \delta_i P_i)},
\ee
and $P_i^0$ is the unique solution to the following equation when $w_i \le 1-\beta(P_i^{\max})$,
\bee
\beta(P_i) = 1 - w_i. \label{equ_pi0}
\eee
\end{pro}
\begin{IEEEproof}
We can prove this proposition by analyzing the first and second derivative of $U_i(P_i)$ with respect to $P_i$. For full proof, please see Appendix A.
\end{IEEEproof}

%Fig.\ref{fig1} is an example for $U_i(\mathbf{P})=U_i(P_i)$. It can be seen that $U_i(P_i)$ is quasi-concave.
%\begin{figure}[htbp]
%\centering
%\includegraphics[width=10cm,draft=false]{u1.eps}
%\caption{Utility v.s. power. $\delta_i = 100$, $w_i = 0.5$, $P_i^c = 0.1$, $P_i^{\max}=1$, $\alpha_i = \frac{1}{SE_{\max}^wEE_{\max}^{1-w}}=0.1331$. }
%   \label{fig1} %% label for entire figure
%\end{figure}

%Note that each user can use Newton-Raphson iteration method to get $P_i^*$ distributed from (\ref{equ1}). Let $\gamma(P_i) = \delta_i(P_i + P_i^c) - (1 -w_i)(1 + \delta_i P_i)\log(1 + \delta_i P_i)$. Then the $(k+1)$-th iteration of optimal power is updated by
%\begin{equation}
%\label{eq:iteration}
%P^u_i(k + 1) = \max\{0, P^u_i(k) - \frac{{\gamma(P^u_i(k))}}{{\gamma'(P^u_i(k))}}\}.
%\end{equation}

Let us denote optimal solution under individual power constraints as $\mathbf{P}^u =  \{P_1^u, P_2^u, \cdots, P_N^u\}$.

Now we consider the power sum constraint in (\ref{equ_cons2}). In $(\mathbf{P1})$, since the feasible region is a compact set and the objective function is continuous, a global optimal solution can be attained. Let us denote the global optimal solution as $\mathbf{P}^* = \{P_1^*, P_2^*, \cdots, P_N^*\}$. Then we have the following proposition.
\begin{pro}
  \label{pro2}
  $\mathbf{P}^* \le \mathbf{P}^u$, i.e.,
  $P_i^* \le P_i^u$, $\forall i \in \{1,2, \cdots, N\}$.
\end{pro}
\begin{IEEEproof}
$\forall i \in \{1,2, \cdots, N\}$, suppose $P_i^* > P_i^u$. Then $\sum_{k \neq i} P_k^*  + P_i^u  < \sum_{k=1}^{N} P_k^*  \le P_{\max}$, which means $\mathbf{P}' = \{P_1^*, \cdots, P_{i-1}^*, P_i^u, P_{i+1}^*, \cdots, P_{N}^*\}$ is a feasible solution to $(\mathbf{P1})$.
According to the \emph{Proposition \ref{pro1}}, we have $U_i(P_i^*) < U_i(P_i^u)$. So $\sum_{k=1}^{N}U_k(P^*_k) < \sum_{k\neq i} U_k(P^*_k) + U_i(P_i^u)$, which is a contradiction to the fact that $\mathbf{P}^*$ is the optimal solution to $(\mathbf{P1})$. This completes the proof.
\end{IEEEproof}

\emph{Proposition 2} shows that for any user $i$, the optimal transmit power $P_i^*$ cannot be greater than $P_i^u$. Therefore we have the following main result of this section.
\begin{theo}
\label{theo1}
$(\mathbf{P1})$ is equivalent to the following problem,
\bee
   \text{maximize} && \sum_{i=1}^{N} U_i(P_i) \nnb \\
(\mathbf{P2}) \quad  \text{subject to}  &&  \label{equ_cons12} 0 \le P_i \le P^{u}_i,  \quad \forall i\\
&& \label{equ_cons22} \sum_{i=1}^{N} P_i \le P_{\max}.
\eee
In addition, $(\mathbf{P2})$ is a convex programming problem.
\end{theo}
\begin{IEEEproof}
Following from \emph{Proposition \ref{pro2}}, we immediately conclude that $(\mathbf{P1})$ is equivalent to $(\mathbf{P2})$.
 In addition, from \emph{Proposition \ref{pro1}}, we know that $U_i(P_i)$ is strictly concave at $P_i \in [0, P_i^u]$.
Thus, ($\mathbf{P2}$) is a problem to maximize a strictly concave function in a convex region, which means it is a convex problem now. This completes the proof.
\end{IEEEproof}
\subsection{Some Analysis}
Next we will give some analysis for the optimal solution $\mathbf{P}^*$ in the following two cases.

\emph{Case 1: $\sum_{i=1}^{N} P_i^u \le P_{\max} $}

In this case, $\mathbf{P}^u$ is feasible for $(\mathbf{P2})$, so it is also the optimal solution for $(\mathbf{P2})$, i.e., $\mathbf{P}^* = \mathbf{P}^u$.

\emph{Case 2: $\sum_{i=1}^{N} P_i^u  >  P_{\max} $}

In this case, we can further narrow down the feasible region for ($\mathbf{P2}$) and achieve the following proposition.
\begin{pro}
  \label{pro3}
If $\sum_{i=1}^{N} P_i^u >  P_{\max} $,  ($\mathbf{P2}$) is equivalent to the following convex optimization problem,
\bee
\text{maximize} && \sum_{i=1}^{N} U_i(P_i) \nnb \\
(\mathbf{P3}) \quad  \text{subject to}  &&  \label{equ_cons13} 0 \le P_i \le P^{u}_i,  \quad \forall i\\
&& \label{equ_cons23} \sum_{i=1}^{N} P_i = P_{\max}.
\eee

\end{pro}
\begin{IEEEproof}
Suppose $\sum_{i=1}^{N} P^*_i < P_{\max}$. Since $\sum_{i=1}^{N} P_i^u  >  P_{\max} $, there exists at least one $i \in \{1,2, \cdots, N\}$ such that $P_i^* < P_i^u$ (Otherwise, $\sum_{i=1}^{N} P^*_i  = \sum_{i=1}^{N} P_i^u  >  P_{\max}$, which is  a contradiction). Therefore, there exists a $\epsilon > 0$ such that $P_i^* + \epsilon \le P_i^u$ and $\sum_{k \neq i} P_k^* + (P_i^* + \epsilon)  \le P_{\max}$. So  $\mathbf{P}' = \{P_1^*, \cdots, P_{i-1}^*, P_i^* + \epsilon, P_{i+1}^*, \cdots, P_{N}^*\}$  is a feasible solution for ($\mathbf{P2}$). According to the \emph{Proposition \ref{pro1}}, we have $U_i(P_i^*) < U_i(P_i^* + \epsilon)$. Then $\sum_{k=1}^{N}U_k(P^*_k) < \sum_{k\neq i} U_k(P^*_k) + U_i(P_i^* + \epsilon)$, which is a contradiction to the fact that $\mathbf{P}^*$ is the optimal solution to $(\mathbf{P1})$. Therefore, we must have $\sum_{i=1}^{N} P^*_i = P_{\max}$, which completes the proof.
\end{IEEEproof}

\subsection{Centralized Algorithm} \label{sec_cenago}
Based on the above analysis, we can readily get the optimal power allocation $\mathbf{P}^*$ for ($\mathbf{P1}$) with a centralized algorithm, as shown in Algorithm \ref{algo1}. In practice, we can implement such centralized algorithm as follows. First, each user $i$ transmits its parameters, including $P_i^{\max}, P_i^c$ and $w_i$ to Node-B. After collecting all the information of all users, Node-B runs Algorithm \ref{algo1} to obtain the optimal power allocation $\mathbf{P}^*$, and then updates the optimal transmit power $P_i^*$ to each user $i$. Finally, each user transmits data at the optimal transmit power.

\begin{algorithm}[ht]
 %\scriptsize
 \caption{Centralized Algorithm for ($\mathbf{P1}$)}
 \label{algo1}
\begin{algorithmic}[1]
        \FOR{$1 \le i \le N$}
            \IF{$w_i > 1 - \beta(P_i^{\max})$}
                \STATE $P_i^u=P_i^{\max}$;
            \ELSE
                \STATE Get $P_i^0$ with Newton-Raphson iteration method for the equation (\ref{equ_pi0});
                \STATE $P_i^u = P_i^0$;
            \ENDIF
        \ENDFOR

        \IF{$\sum_{i=1}^{N} P_i^u \le P_{\max} $}
            \STATE $\mathbf{P}^*=\mathbf{P}^u$;
        \ELSE
            \STATE Get $\mathbf{P}^*$ with gradient projection method for ($\mathbf{P3}$);
        \ENDIF
\end{algorithmic}
\end{algorithm}

\section{Distributed Primal-Dual Implementation}
In the previous section, we provide Algorithm \ref{algo1} to solve the utility maximization problem in a centralized manner. However, it requires Node-B to have knowledge of all the global information of all the users. Furthermore, the centralized algorithm still incurs some computational complexity and is not robust against temporary variation of system parameters, such as instantaneous CSI. Hence, we hope to implement the algorithm in a distributed manner. Inspired by the distributed algorithm in network flow optimization problem \cite{SR07}, we design the following primal-dual distributed algorithm to achieve the optimal power allocation $\mathbf{P}^*$,
\bee
\label{equ_dis}
\begin{cases}
\dot{P_i}= k_i[U'_i(P_i) - \lambda]_{P_i}^{P_i^u+}, \forall i \in \{1,2,\cdots, N\}\\
\dot{\lambda} = g[\sum_{i=1}^{N} P_i - P_{\max}]_{\lambda}^{+},
\end{cases}
\eee
where
\be
[f]_{z}^{+} =
\begin{cases}
\max(f, 0), & z \le 0 \\
f,  &  z > 0
\end{cases}
\ee
and
\be
[f]_{z}^{a+} =
\begin{cases}
\max(f, 0), & z \le 0 \\
\min(f, 0), & z \ge a \\
f,  &  0 < z < a
\end{cases}
\ee
and $k_i$ and $g$ are positive stepsize.

From (\ref{equ_dis}), each user does not need the information of others but just the penalty $\lambda$. The only overhead is that Node-B broadcasts $\lambda$ to all users and each user $i$ updates $P_i$ to Node-B until convergence. Therefore, such implementation only consumes a small amount of overhead between each user and Node-B. Also, our proposed distributed algorithm reduces computational complexity compared to Algorithm \ref{algo1}. Moreover, we further prove that this primal-dual distributed algorithm in (\ref{equ_dis}) can converge to the global optimal, as shown in the following theorem.
\begin{theo}
\label{theo2}
The distributed algorithm in (\ref{equ_dis}) is globally asymptotically stable and the only equilibrium is $\mathbf{P}^*$.
\end{theo}
\begin{IEEEproof}
We can prove this theorem by constructing the following Lyapunov function,
\be
V(\mathbf{P}, \lambda) = \frac{1}{2}\sum_{i=1}^N \frac{(P_i-P_i^*)^2}{k_i} + \frac{(\lambda-\lambda^*)^2}{2g},
\ee
where $(P_1^*, P_2^*, \cdots, P_N^*, \lambda^*)$ satisfy the KKT conditions for $(\mathbf{P2})$ and $\lambda^*$ is the multiplier for (\ref{equ_cons22}).
For full proof, please see Appendix B.
\end{IEEEproof}

%\textbf{\textit{Remark:}} In this paper, Node-B can be regarded as a central controller for all users, so Node-B can get global information for all users and run the centralized algorithm, as discussed in Section \ref{sec_cenago}. From this angle, our primal-dual distributed algorithm may not outperform the centralized algorithm. This is because we only consider static parameters including CSI in this paper so that Node-B just need one time to collect all information. However, when we consider time-varying system, we believe our primal-dual distributed algorithm can not only reduce overhead but also be robust against instantaneous CSI and other time-varying parameters. This will be one of our further topics.

\section{Numerical Results}
In this section, simulation results are provided to validate our theoretic analysis, which show that our approach not only captures user diversity for EE-SE tradeoff, but also ensures user fairness. In addition, we also verify that our primal-dual distributed algorithm can converge to the global optimal solution.
Throughout this section, we will set the maximal transmit power to be $1$W ($30$dBm) for all users, by adopting the transmitter's power level $1$ for 1800/1900 MHz mobile phones \cite{3GPP-TS}. The circuit power is set to be $0.1$W for all users.

\subsection{User Diversity for EE-SE Tradeoff}
As shown in (\ref{equ_uui}), different users can have different preferences for SE and EE according to $w_i$.
In this part, we will show how to capture user diversity with $w_i$. We consider two users, i.e., $N=2$, and let the number of receiving antennas for Node-B to be $2$, i.e., $M=2$. In addition, we regard two users as homogeneous such that they have the same CSI. Specifically, let $\delta_1=\delta_2=20$dB and $P_{\max}=1.5$W.

\begin{figure*}
\hspace{-0.5cm}
  \subfigure[Optimal transmit power $P_1^*$]{
  \begin{minipage}[t]{0.33\linewidth}
    \centering
    \includegraphics[width=\linewidth]{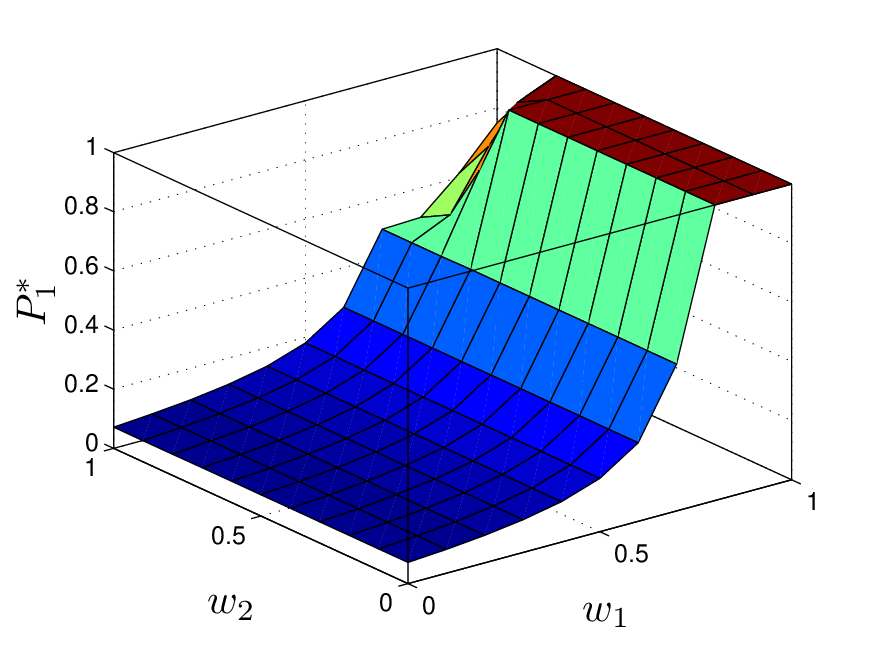}\\
     \label{fig_user_diversity_P1}
  \end{minipage}}
  \subfigure[$SE_1$ with $P_1^*$]{
  \begin{minipage}[t]{0.33\linewidth}
    \centering
    \includegraphics[width=\linewidth]{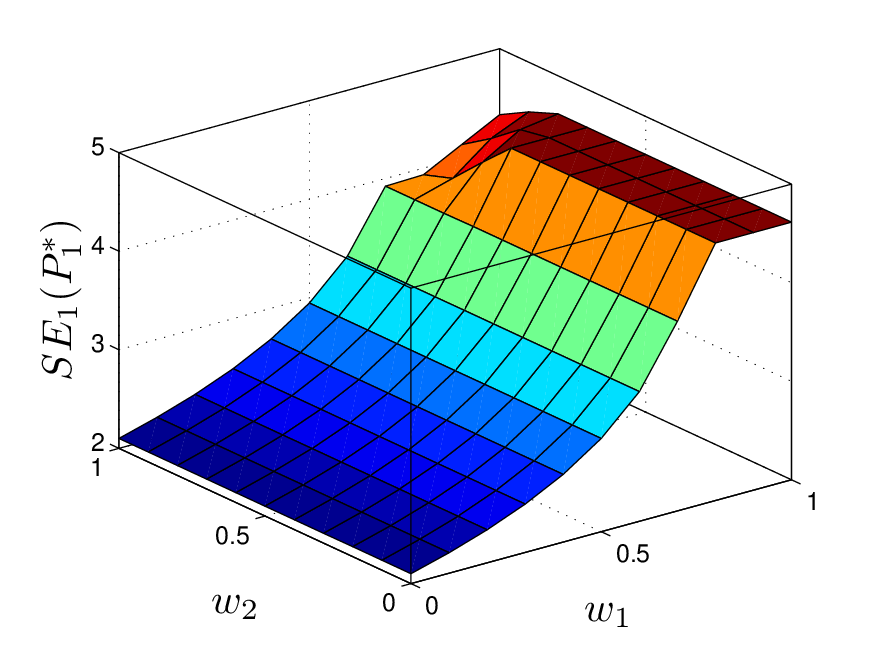}\\
        \label{fig_user_diversity_SE1}
  \end{minipage}}
  \subfigure[$EE_1$ with $P_1^*$]{
  \begin{minipage}[t]{0.33\linewidth}
    \centering
    \includegraphics[width=\linewidth]{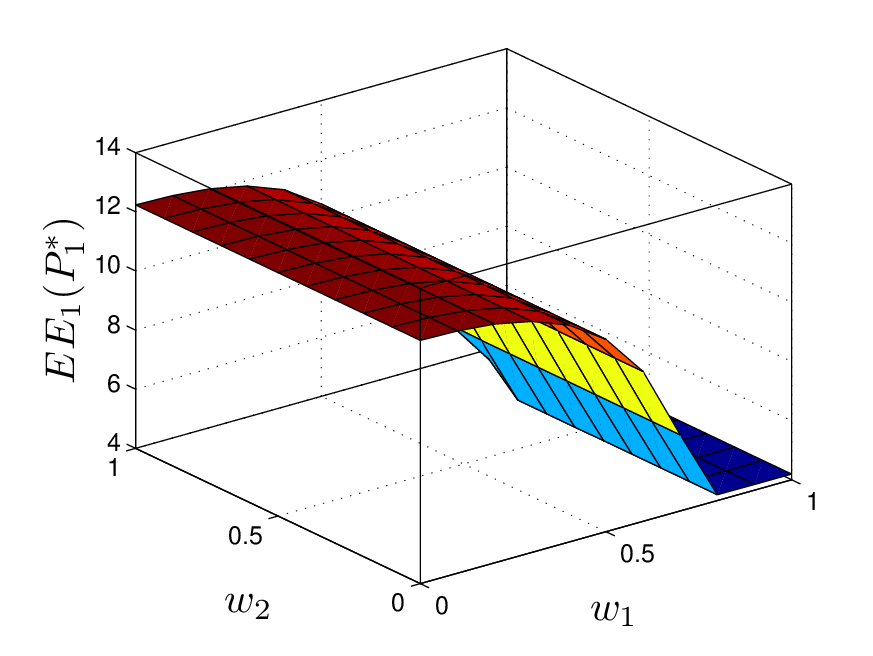}\\
        \label{fig_user_diversity_EE1}
 \end{minipage}}%
 \\
   \subfigure[Optimal transmit power $P_2^*$]{
  \begin{minipage}[t]{0.33\linewidth}
    \centering
    \includegraphics[width=\linewidth]{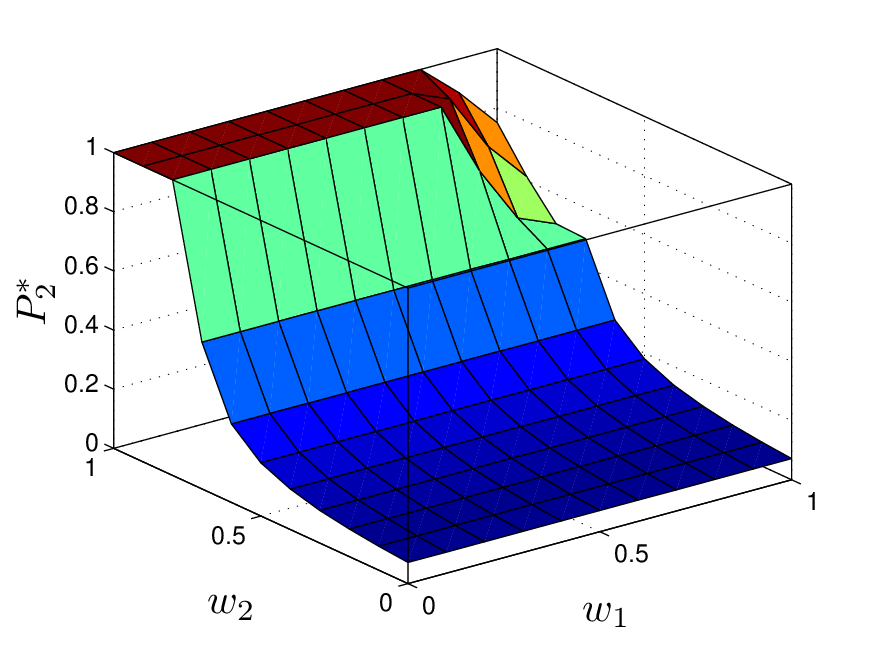}\\
        \label{fig_user_diversity_P2}
  \end{minipage}}
  \subfigure[$SE_2$ with $P_2^*$]{
  \begin{minipage}[t]{0.33\linewidth}
    \centering
    \includegraphics[width=\linewidth]{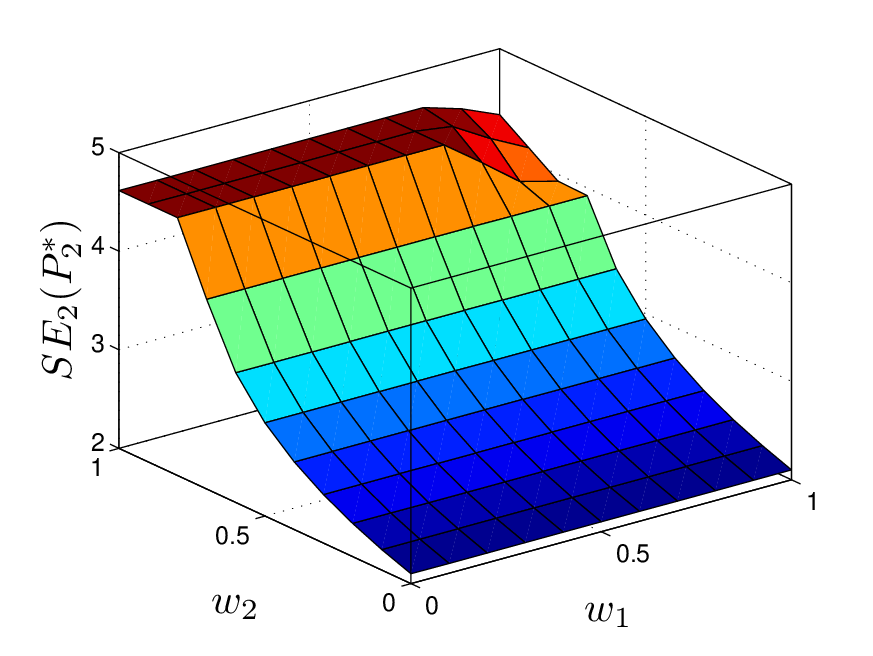}\\
        \label{fig_user_diversity_SE2}
  \end{minipage}}
  \subfigure[$EE_2$ with $P_2^*$]{
  \begin{minipage}[t]{0.33\linewidth}
    \centering
    \includegraphics[width=\linewidth]{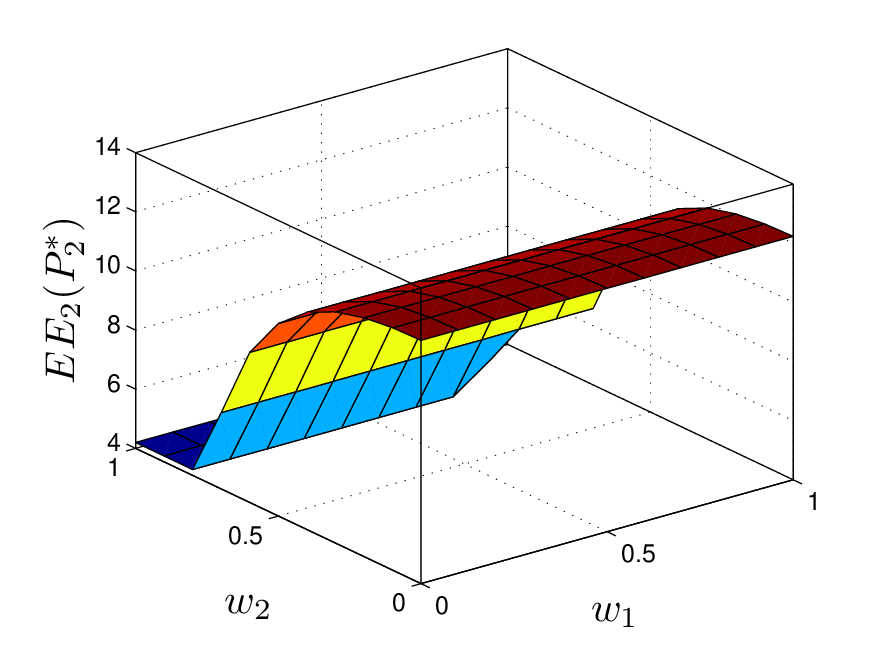}\\
        \label{fig_user_diversity_EE2}
 \end{minipage}}%
 \caption{User diversity for EE-SE tradeoff with $N=2, M=2, P_1^{\max}=P_2^{\max}=1\text{W}, P_{\max}=1.5\text{W}, P_1^c=P_2^c=0.1\text{W}, \delta_1=\delta_2=20\text{dB}$.}
 \label{fig_user_diversity}
\end{figure*}

Fig. \ref{fig_user_diversity} shows the impact of users' different preferences, i.e., $w_1$ and $w_2$. From Fig. \ref{fig_user_diversity_P1} and \ref{fig_user_diversity_P2}, we can see the optimal transmit power for two users. Under the optimal transmit power for user $1$, Fig. \ref{fig_user_diversity_SE1} demonstrates that SE increases as $w_1$ increases while Fig. \ref{fig_user_diversity_EE1} demonstrates that EE decreases as $w_1$ increases. Similarly, Fig. \ref{fig_user_diversity_SE2} and \ref{fig_user_diversity_EE2} show the same effect of $w_2$ for user $2$. Such results comply with our intuition for the effect of $w_i$ in (\ref{equ_uui}), and therefore verify that our utility function can capture user diversity for EE-SE tradeoff very well with the preference $w_i$.

\subsection{User Fairness}
In this part, we will show that our utility function in (\ref{equ_ui}) can ensure fairness among all users.
We still consider two users, i.e., $N=2$, and let $M=2$. We set $w_1=w_2=0.5$ which means SE and EE are equally important for both users. $P_{\max}$ is set to be $1.5$W. We will vary the channel conditions for two users. Specifically, we fix $\delta_1$ to three different levels: $-20$dB (worst), $0$dB (normal) and $20$dB (best) respectively, while changing $\delta_2$ from $-20$dB (worst) to $20$dB (best). In addition, we will use the Jain's fairness index \cite{JC84} to evaluate the user fairness, i.e.,
\be
\text{Jain's Fairness Index} = \frac{[\exp(U_1) + \exp(U_2)]^2}{2[(\exp(U_1))^2+(\exp(U_2))^2]},
\ee
where $U_1$ and $U_2$ are the utility from \eqref{equ_ui} for user 1 and user 2 with channel conditions $\delta_1$ and $\delta_2$, respectively.

\begin{figure}
  \centering
  % Requires \usepackage{graphicx}
  \includegraphics[width=\linewidth]{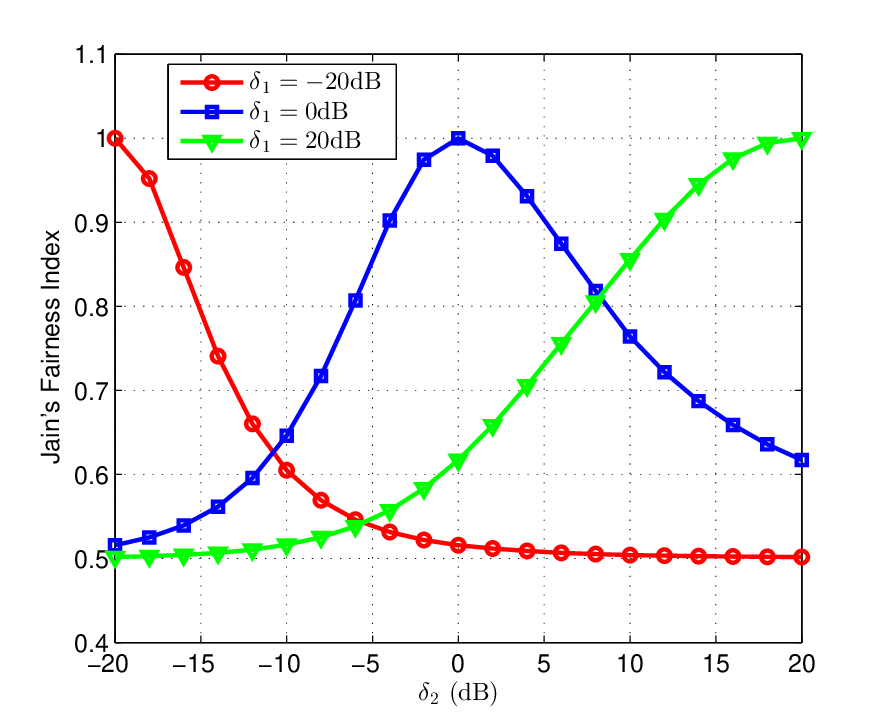}\\
  \caption{User fairness with $N=2, M=2, P_1^{\max}=P_2^{\max}=1\text{W}, P_{\max}=1.5\text{W}, P_1^c=P_2^c=0.1\text{W}, w_1=w_2=0.5$.}\label{fig_Jain_fairness}
\end{figure}

Fig. \ref{fig_Jain_fairness} shows that the closer the channel conditions are, the better the fairness is.
In addition, when $\delta_1=\delta_2$, the index is 1 (the best fairness) which means no bias exists and two users have the same utility. Furthermore, even though when the channel condition is worst for user $1$ with $\delta_1=-20$dB and the channel condition is best for user $2$ with $\delta_2=20$dB, the index still does not touch 0.5 (the worst fairness) exactly and it is actually 0.5017. This  means user $1$ can still transmit data with a positive transmit power.
Therefore, user fairness can be guaranteed under our proposed utility function in (\ref{equ_ui}).

\subsection{Primal-Dual Distributed Algorithm}
In this part, we will validate the effectiveness of our primal-dual distributed algorithm in (\ref{equ_dis}). We consider four users, i.e., $N=4$ and let $M=4$. We set different EE-SE preferences for all users, which are $w_1=0, w_2=0.3, w_3=0.7$ and $w_4=1$. The maximal sum power is $P_{\max}=3$W. In addition, the step sizes, $k_i$, $\forall i \in \{1,2,3,4\}$ and $g$ in (\ref{equ_dis}) are set to be $0.001$.

\begin{figure}[htbp]
    \subfigure[Power Convergence]{
        \label{fig_power_convergence}
        \begin{minipage}[b]{\linewidth}
          \centering
          \includegraphics[width=\linewidth]{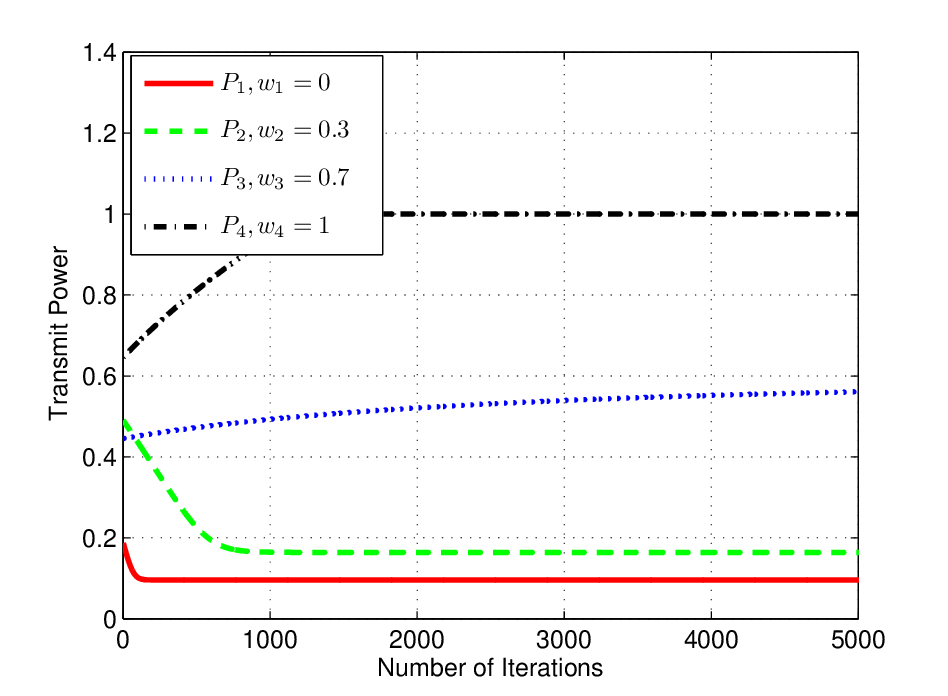}
        \end{minipage}}%
        \\
    \subfigure[Utility convergence]{
        \label{fig_utility_convergence}
        \begin{minipage}[b]{\linewidth}
          \centering
          \includegraphics[width=\linewidth]{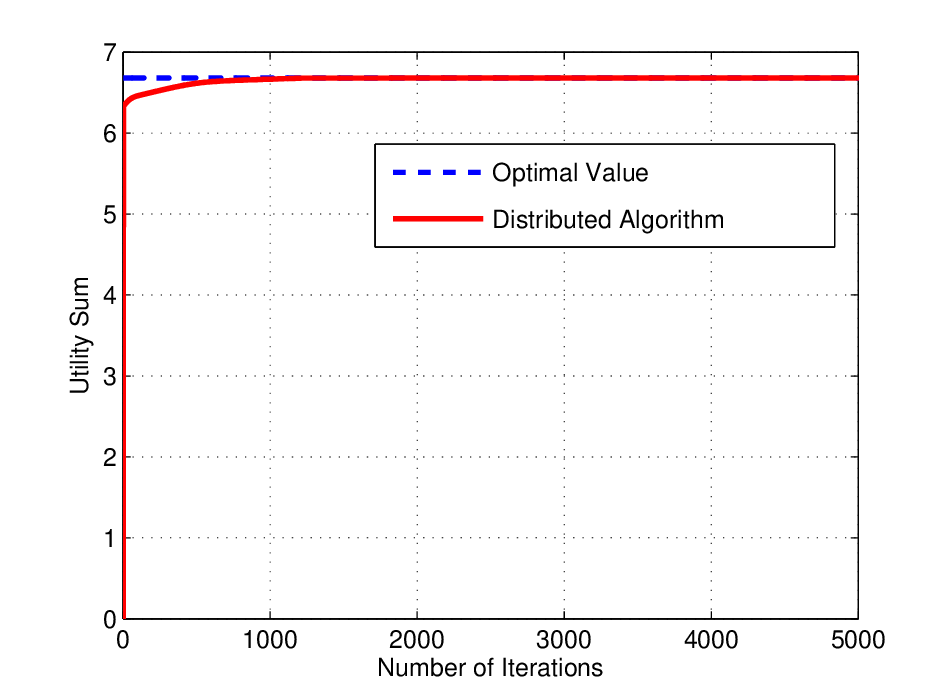}
        \end{minipage}}%
        \caption{Convergence of primal-dual distributed algorithm.}
\label{fig_convergence} %% label for entire figure
\end{figure}

Fig. \ref{fig_convergence} shows the simulation results. From Fig. \ref{fig_power_convergence}, we can see that the power allocation can be converged and it also shows that our approach can capture user diversity when different users have different preferences for EE-SE tradeoff. From Fig. \ref{fig_utility_convergence}, we can see that the distributed algorithm can converge to the global optimum, which verifies \emph{Theorem \ref{theo2}}.

\section{Conclusion and Future Work}
In this paper, we consider utility maximization for the uplink MU-MIMO system. We define the utility function combining both EE-SE tradeoff and user fairness. After formulating the utility maximization problem with individual power constraints and sum power constraint, we analyze the optimal power allocation scheme. Although the original optimization problem is not convex, we propose a convexization procedure to convert it into an equivalent convex programming problem, which has been proven to have the same global optimal solution as the original problem. Moreover, we have proposed two algorithms to obtain the optimal solution: one is the centralized algorithm which requires knowledge of all the global information; the other is the primal-dual distributed algorithm which only needs a small amount of overhead between each user and Node-B. Furthermore, we have proved that our proposed distributed algorithm can converge to the global optimal solution.

Several extensions can be done in the future to apply our results to more practical scenarios. First, if we use other detection methods rather than ZF, such as minimum mean-squared error (MMSE), the problem becomes more complex because of the coupled interference part. Second, static CSI is investigated in this paper, but more realistic version should be stochastic CSI under which we should analyze the achievable (or average) SE and EE to guarantee a statistical long-term performance. Last but not least, user pairing, which depicts how to select part of users from all available users to form the MU-MIMO system, is also an important issue in uplink MU-MIMO system. Traditional metric is to maximize throughput or SE, but the story will be changed when we use EE-SE tradeoff as the new metric.
In summary, this paper opens a new way to \textit{quantitatively} analyze the EE-SE tradeoff with fairness guarantee for MU-MIMO system.

\vspace{0.2in}
\appendices

\section{Proof of Proposition 1}
The first derivative of $U_i(P_i)$ in ($\ref{equ_ui}$) is
\bee
U'_i(P_i) &=& \frac{1}{\log(1+\delta_i P_i)}\cdot\frac{\delta_i}{1+\delta_i P_i}-\frac{1-w_i}{P_i+P_i^c} \nnb \\
&=& \frac{1}{P_i + P_i^c}[\frac{\delta_i(P_i+P_i^c)}{(1+\delta_i P_i)\log(1+\delta_i P_i)}-(1-w_i)] \nnb \\
&=& \frac{1}{P_i + P_i^c}[\beta(P_i)-(1-w_i)],
\eee
where
\be
\label{beta}
\beta(P_i)=\frac{\delta_i(P_i + P_i^c)}{(1 + \delta_i P_i)\log(1 + \delta_i P_i)}.
\ee
And the second derivative of $U_i(P_i)$ is
\be
U''_i(P_i) = \frac{\beta'(P_i)(P_i + P_i^c) - [\beta(P_i) - (1-w_i)]}{(P_i + P_i^c)^2}.
\ee

In (\ref{beta}), we can get first derivative of $\beta(P_i)$ as
\bee
&&\beta'(P_i) \nnb \\
&& = \delta_i \cdot \frac{[\log(1 + \delta_i P_i)-\delta_i P_i] - P_i^c \delta_i [\log(1+\delta_i P_i) + 1]}{[(1 + \delta_i P_i)\log(1 + \delta_i P_i)]^2}  \nnb \\
&& < 0,
\eee
which shows that  $\beta(P_i)$ is strictly decreasing with $P_i$.

\emph{Case 1:} If $w_i > 1 - \beta(P_i^{\max})$, then $1-w_i < \beta(P_i^{\max}) \le \beta(P)$, which yields $U'_i(P_i) > 0$. So $U_i(P_i)$ is strictly increasing at $[0, P_i^{\max}]$. The utility $U_i(P_i)$ is maximized when $P_i = P_i^{\max}$. In addition, $U''_i(P_i) < 0$, so $U_i(P_i)$ is strictly concave at $[0, P_i^{\max}]$.

\emph{Case 2:} If $w_i \le 1 - \beta(P_i^{\max})$, then $U'_i(P_i) = 0$ has one and only one solution $P_i^0 \in (0, P_i^{\max}]$. This is because $\beta(P_i)$ is strictly decreasing and $\lim\limits_{P_i \rightarrow 0} \beta(P_i) = +\infty$ and $\beta(P_i^{\max}) \le 1-w_i$. Note that $P_i^0$ is the unique solution of the following equation,
\be
\beta(P_i)=\frac{\delta_i(P_i + P_i^c)}{(1 + \delta_i P_i)\log(1 + \delta_i P_i)} = 1 -w_i.
\ee
At the interval $[0, P_i^u)$, we have $\beta(P_i) - (1-w_i) > 0$, so $U'_i(P_i) > 0$ and $U''_i(P_i) < 0$. At the interval $[P_i^0, P_i^{\max}]$, $\beta(P_i) - (1-w_i) \le 0$, so $U'_i(P_i) \le 0$. Then we obtain that the utility $U_i(P_i)$ is maximized at $P=P_i^0$ and $U_i(P_i)$ is strictly increasing and concave at $[0, P_i^0]$ and strictly decreasing at $(P_i^0, P_i^{\max}]$.

The proof is completed.

\section{Proof of Theorem 2}
\begin{IEEEproof}
The proof is based on \cite{Wen04}. Readers can find more information about primal-dual algorithm for network flow optimization in \cite{SR07} \cite{Wen04}.

First, we rewrite the problem $(\mathbf{P2})$,
\bee
\text{maximize} && \sum_{i=1}^{N} U_i(P_i) \nnb \\
\text{subject to}  &&  \label{equ_cons14} -P_i \le  0, \quad \forall i\\
&&  \label{equ_cons24} P_i - P_i^u \le 0 , \quad \forall i\\
&&  \label{equ_cons34} \sum_{i=1}^{N} P_i - P_{\max} \le 0.
\eee
Since $\mathbf{P}^*=\{P_1^*, P_2^*, \cdots, P_N^*\}$ is the optimal solution of this problem and the Slater's condition holds for this convex problem, there exists KKT multipliers $\mu_i^*$ for (\ref{equ_cons14}), $\nu_i^*$ for (\ref{equ_cons24}) and $\lambda^*$ for (\ref{equ_cons34}) such that they satisfy the following KKT conditions,
\bee
&& U'_i(P_i^*)+\mu^*_i-\nu^*_i - \lambda^* = 0, \forall i \label{equ_KKT1}\\
&& 0 \le P_i^* \le P_i^u, \forall i \label{equ_KKT2}\\
&& \sum_{i=1}^{N}P_i^* \le P_{\max}, \label{equ_KKT3}\\
&& \mu^*_i \ge 0, \nu^*_i \ge 0, \lambda^* \ge 0, \forall i \label{equ_KKT4}\\
&& \mu^*_i P_i^* = 0, \forall i \label{equ_KKT5}\\
&& \nu^*_i (P_i^* - P_i^u) = 0, \forall i \label{equ_KKT6}\\
&& \lambda^*(\sum_{i=1}^{N}P_i^* - P_{\max}) = 0. \label{equ_KKT7}
\eee
Now given such KKT conditions,  we can prove this theorem by constructing the following Lyapunov function for the system in (\ref{equ_dis}),
\be
V(\mathbf{P}, \lambda) = \frac{1}{2}\sum_{i=1}^N \frac{(P_i-P_i^*)^2}{k_i} + \frac{(\lambda-\lambda^*)^2}{2g}. \label{equ_lya}
\ee
It is easy to verify that $V(\mathbf{P}, \lambda)$ in (\ref{equ_lya}) is positive definite. To show that $V(\mathbf{P}, \lambda)$ is a Lyapunov function, it suffices to verify its Lie derivative with respect to (\ref{equ_dis}) is nonnegative, i.e., $\dot{V} \le 0 $. This is true because,
\bee
\dot{V} &=& \sum_{i=1}^{N}\frac{P_i-P_i^*}{k_i}\dot{P_i} + \frac{\lambda-\lambda^*}{g} \dot{\lambda} \\
&=& \sum_{i=1}^{N}(P_i-P_i^*)[U'_i(P_i) - \lambda]_{P_i}^{P_i^u+}  \nnb \\
&& + (\lambda-\lambda^*) [\sum_{i=1}^{N} P_i - P_{\max}]_{\lambda}^{+}  \\
&\le & \sum_{i=1}^{N}(P_i-P_i^*)[U'_i(P_i) - \lambda] \nnb  \\
&& + (\lambda-\lambda^*) [\sum_{i=1}^{N} P_i - P_{\max}] \\
%&=&  \sum_{i=1}^{N}(P_i-P_i^*)[U'_i(P_i) - U'_i(P_i^*) + U'_i(P_i^*) - \lambda^* + \lambda^* - \lambda] \nnb \\
%&& + (\lambda-\lambda^*) [\sum_{i=1}^{N} P_i - P_{\max}]\\
&=& \sum_{i=1}^{N}(P_i-P_i^*)[U'_i(P_i) - U'_i(P_i^*)] \label{equ_negative1}\\
&& + \sum_{i=1}^{N}(P_i-P_i^*)[U'_i(P_i^*) - \lambda^*] \label{equ_negative2}\\
&& + (\lambda-\lambda^*) [\sum_{i=1}^{N} P_i^* - P_{\max}] \label{equ_negative3}\\
&\le& 0,
\eee
where (\ref{equ_negative1}) is nonpositive since $U'(P_i)$ is strictly decrease over the interval $[0, P_i^u]$ according to \textit{Proposition 1},
and (\ref{equ_negative3}) is nonpositive following from the KKT conditions as,
\bee
&& (\lambda-\lambda^*) [\sum_{i=1}^{N} P_i^* - P_{\max}] \nnb \\
&=& \lambda [\sum_{i=1}^{N} P_i^* - P_{\max}] - \lambda^* [\sum_{i=1}^{N} P_i^* - P_{\max}] \nnb \\
&=& \lambda [\sum_{i=1}^{N} P_i^* - P_{\max}], \quad \//\// \text{from (\ref{equ_KKT7})} \\
&\le& 0, \quad \//\// \text{from (\ref{equ_KKT4})}
\eee
and (\ref{equ_negative2}) is nonpositive following from the KKT conditions as,
\bee
&& \sum_{i=1}^{N}(P_i-P_i^*)[U'_i(P_i^*) - \lambda^*] \nnb \\
&=& \sum_{i=1}^{N}(P_i-P_i^*)(\nu^*_i - \mu^*_i ) \quad \//\// \text{from (\ref{equ_KKT1})}  \\
&=& \sum_{i=1}^{N}(P_i-P_i^*)\nu^*_i - \sum_{i=1}^{N}(P_i-P_i^*)\mu^*_i \\
&=& \sum_{i=1}^{N}(P_i-P_i^*)\nu^*_i - \sum_{i=1}^{N}P_i\mu^*_i + \sum_{i=1}^{N}P_i^*\mu^*_i \\
&\le& \sum_{i=1}^{N}(P_i-P_i^*)\nu^*_i \quad \//\// \text{from (\ref{equ_KKT2}, \ref{equ_KKT4}, \ref{equ_KKT5})} \\
&=& \sum_{i=1}^{N}(P_i-P_i^u)\nu^*_i + \sum_{i=1}^{N}(P_i^u-P_i^*)\nu^*_i \\
&=& \sum_{i=1}^{N}(P_i-P_i^u)\nu^*_i  \quad \//\// \text{from (\ref{equ_KKT6})} \\
&\le& 0. \quad \//\// \text{from (\ref{equ_KKT2}, \ref{equ_KKT4})}
\eee
Therefore, we have verified $V(\mathbf{P}, \lambda)$ in (\ref{equ_lya}) is a Layponouv function. So the primal-dual distributed system in (\ref{equ_dis}) is globally asymptotically stable and will converge to the equilibria set $\{\mathbf{P}: \dot{V}=0\}$ \cite{Khalil02}. On the other hand, $\dot{V}=0$ only if (\ref{equ_negative1}) is equal to 0, i.e.,
\be
\sum_{i=1}^{N}(P_i-P_i^*)[U'_i(P_i) - U'_i(P_i^*)] = 0,
\ee
which holds only if $P_i=P_i^*$ for all $i \in \{1,2,\cdots, N\}$. Therefore, the equilibria set $\{\mathbf{P}: \dot{V}=0\}$ only contains one point, i.e., $\mathbf{P}^*$. This completes the proof.
\end{IEEEproof}

% that's all folks
\end{document}